\documentclass[aps,twocolumn,prl,superscriptaddress,showpacs,amsmath,amssymb]{revtex4}
\usepackage{amsthm,graphicx}

\newcommand{\bra}[1]{\mathinner{\langle{#1}|}}
\newcommand{\ket}[1]{\mathinner{|{#1}\rangle}}

\newcommand{\proj}[1]{\mathinner{|{#1}\rangle\langle{#1}|}}

\DeclareMathOperator{\tr}{tr} 

\begin{document}

\title{Broadcasting Quantum Fisher Information}

\author{Xiao-Ming Lu} \email{luxiaoming@gmail.com} \affiliation{Centre for Quantum Technologies, National University of
  Singapore, 3 Science Drive 2, Singapore 117543, Singapore}

\author{Zhe Sun} \affiliation{Department of Physics, Hangzhou Normal
  University, Hangzhou 310036, China}

\author{Xiaoguang Wang} \affiliation{Zhejiang Institute of Modern Physics and
  Department of Physics, Zhejiang University, Hangzhou 310027, China}

\author{Shunlong Luo}\email{luosl@amt.ac.cn} \affiliation{Academy of Mathematics and Systems Science,
  Chinese Academy of Science, Beijing 100190, China}

\author{C. H. Oh} \email{phyohch@nus.edu.sg} \affiliation{Centre for Quantum Technologies, National University of
  Singapore, 3 Science Drive 2, Singapore 117543, Singapore}
\affiliation{Department of Physics, National University of Singapore, 3
  Science Drive 2, Singapore 117543, Singapore}

\begin{abstract}
  It is well known that classical information can be cloned, but
  non-orthogonal quantum states cannot be cloned, and non-commuting quantum
  states cannot be broadcast. We conceive a scenario in which the object we
  want to broadcast is the statistical distinguishability, as quantified by
  quantum Fisher information, about a signal parameter encoded in quantum
  states. We show that quantum Fisher information cannot be cloned, whilst it
  might be broadcast even when the input states are non-commuting. This
  situation interpolates between cloning of classical information and
  no-broadcasting of quantum information, and indicates a hybrid way of
  information broadcasting which is of particular significance from both practical and theoretical
  perspectives.
\end{abstract}


\maketitle

One of the most fundamental information tasks in communication is the
dissemination of resources, such as quantum
states~\cite{Wootter1982,Barnum1996,Buzek,
  Mor1998,Koashi1998,Cerf,Scarani2005}, or
correlations~\cite{Horodecki2006,Janzing2007,Piani2008,Piani2009,Luo2010b}.
In contrast to the classical regime, quantum mechanics usually
imposes strong limitations on such a task. Celebrated examples are
no-cloning of non-orthogonal quantum states ~\cite{Wootter1982},
no-broadcasting of non-commuting quantum states ~\cite{Barnum1996},
and no-local-broadcasting of quantum correlations ~\cite{Piani2008,
Luo2010b}. However, in many theoretical and practical issues, it is
not the states, but rather the information about a signal parameter
encoded in the quantum states, that is needed to be disseminated.
The information carried by a physical parameter is usually
synthesized by quantum Fisher information (QFI)~\cite{Helstrom1976,
  Holevo1982, Wootters1981, Braunstein1994, Fuchs1995, Barndorff-Nielsen2000,
  QF0, QF1, QF2, QF3}, which is the minimum achievable statistical uncertainty
in the estimation of the parameter, and plays a fundamental and crucial role
in both quantum foundation and quantum practice such as quantum
metrology~\cite{QM1}. This motivates us to study the cloning and broadcasting
of QFI.

In this Letter, we show that QFI cannot be cloned, whilst it might be
broadcast even when the underlying states encoding the parameter are
non-commuting. In particular, we prove that QFI can be infinitely broadcast if
and only if it is uniform, whose exact meaning will be given later. Moreover,
we identify all states arising from broadcasting of QFI. As a hybrid object
lying between purely classical information and fully quantum information, QFI
is of unique significance in quantum information processing. Our results shed
new insights into the nature of QFI.

\begin{figure} \center{
    \includegraphics[height=3cm]{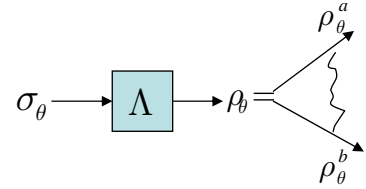}
    \caption {\label{fig:1} {\small The states $\sigma _\theta$ are
        ``broadcast",  by the channel $\Lambda$,  to the bipartite states
        $\rho _\theta$, with reduced states $\rho^a_\theta = \tr_b\rho_\theta, \
        \rho^b_\theta =\tr_a\rho_\theta.$  While
        broadcasting states requires $\rho ^a_\theta =\rho ^b_\theta =\sigma
        _\theta,  \forall \ \theta$,  broadcasting QFI only requires $F(\rho
        ^a_\theta )=F(\rho ^b_\theta )=F(\sigma _\theta), \ \forall \
        \theta.$ Apparently, the former implies the latter, but the converse
        is not true in general. }}}
\end{figure}

{\sl Quantum Fisher information.}--To estimate a physical parameter encoded in
a family of quantum states $\{\rho_\theta\}_{\theta \in \Theta},$ one performs
a measurement $M=\{M_j\},$ mathematically described by a
positive-operator-valued measure, on the states and then forms an estimator
$\hat \theta$ of the parameter in terms of the measurement outcomes. The
measurement $M$ induces a parametric probability distribution
$p_\theta(j):=\tr\rho_\theta M_j$ with the corresponding (classical) Fisher
information $ F(\rho_\theta |M) := \sum_j p_\theta(j) (\partial_\theta {\rm
  log} p_\theta(j))^2 .$ For an unbiased estimator $\hat \theta,$ the
celebrated Cram\'{e}r-Rao inequality~\cite{Helstrom1976,Holevo1982}, $ \Delta
\hat{\theta}:= \langle (\hat{\theta}-\theta) ^2\rangle \geq 1/F(\rho_{\theta}
| M), $ provides a fundamental limitation to the estimation precision. In
order to achieve the best precision, one needs to maximize the Fisher
information $F(\rho _\theta |M)$ over all measurements $M$. The maximum value
is given by QFI~\cite{Braunstein1994}, $F(\rho_\theta) :=\tr \rho_\theta
L_\theta^2=\max _M F(\rho _\theta |M),$ which is an intrinsic measure for the
statistical distinguishability of states of the family $\{\rho _\theta\}$ in
the neighborhood of $\rho_\theta$. Here the symmetric logarithmic derivative
(SLD) $L_\theta$ is determined by the equation $\partial_\theta \rho_\theta =
(L_\theta \rho_\theta+\rho_\theta L_\theta)/2.$ A measurement, which usually
depends on the parameter, achieving the maximum Fisher information is called
an optimal distinguishing measurement. Moreover, QFI is the infinitesimal
metric of both the Bures distance $d_\mathrm{B}(\rho_1,\rho_2 ) = \sqrt{2-2A
  (\rho_1,\rho_2)}$~\cite{Bures1969}, and the statistical
distance~\cite{Wootters1981,Braunstein1994,Fuchs1995}
\begin{equation}
  \label{eq:statistical_distance}
  d_{\rm S}(\rho_1,\rho_2) := \max_ M \arccos\sum_j
  \sqrt{\tr \rho_1 M_j \cdot \tr \rho_2 M_j}
\end{equation}
with which a Riemannian geometry can be endowed on the space of
states \cite{Petz1996,Uhlmann1986}. Here $A (\rho_1,\rho_2):= \tr
(\rho_1^{1/2}\rho_2\rho_1^{1/2})^{1/2}= \cos d_{\rm S}(\rho _1, \rho
_2)$ is the fidelity ~\cite{Jozsa1994}. Consequently, for two
neighboring states $\rho_\theta$ and $\rho_{\theta+d\theta}$, one
has $d_{\rm  B}(\rho_\theta,\rho_{\theta+d\theta}) = d_{\rm  S}
(\rho_\theta,\rho_{\theta+d\theta}) = \frac{1}{4}
F(\rho_\theta)d\theta^2.$ A measurement achieving the equality in
Eq.~(\ref{eq:statistical_distance}) is called an optimal
distinguishing measurement for $\rho _1$ and $\rho_2$.

The task we consider is to disseminate the statistical distinguishability, as
quantified by QFI, of a parameter $\theta\in \Theta$. The setup, as depicted
in Fig. 1, is similar to the conventional cloning and broadcasting scenario:
The states are ``broadcast" to two parties $a$ and $b$ through a channel $
\Lambda : \sigma_\theta \mapsto \rho_\theta \in S(H^a\otimes H^b) $ (state
space). The reduced states for parties $a$ and $b$ are
$\rho_\theta^a=\tr_b\rho_\theta, $ $\rho_\theta^b=\tr_a \rho_\theta,$
respectively. Unlike the conventional broadcasting, we are not interested in
the quantum states themselves, but rather the physical parameter $\theta$. We
say that QFI of $\{\sigma_{\theta}\}_{\theta \in \Theta}$ can be cloned if
there exists a channel $ \Lambda : \sigma_\theta \mapsto \rho_\theta \in
S(H^a\otimes H^b) $ such that (i) $F(\rho_\theta^a) = F(\rho_\theta^b) =
F(\sigma_\theta)$; (ii) $\rho_{\theta}= \rho_\theta^a \otimes \rho_\theta^b$.
We say that QFI can be broadcast if only (i) is required.

 {\sl No-cloning of QFI.}--First, we show that QFI cannot be cloned except for
 the trivial case of vanishing QFI. In fact, there exists a strong restriction
 to the distribution of QFI in case of factorized output states.

 {\bf Theorem} 1. {\sl QFI cannot be cloned in the sense that for any channel
   $ \Lambda \colon \sigma_\theta \mapsto \rho_\theta^a \otimes \rho_\theta^b
   $, if $F(\rho_\theta^{a})=F(\sigma_\theta),$ then $F(\rho_\theta^{b})=0$.
   That is, if one party obtains the same QFI as the input states, then the
   other party cannot have any positive QFI.}

 This is a consequence of the additivity of QFI for factorized states, $
 F(\rho_{\theta}^{a} \otimes \rho_{\theta}^{b}) = F(\rho_{\theta}^{a}) +
 F(\rho_{\theta}^{b}), $ and the monotonicity of QFI under quantum channels, $
 F(\rho_{\theta}^{a} \otimes \rho_{\theta}^{b})\leq F(\sigma _\theta).$ Note
 that $ F(\rho_\theta^b) =0$ if and only if $L_\theta^b
 \sqrt{\rho_\theta^b}=0$, which in turn implies that
 $\partial_\theta\rho_\theta^b = 0$. Here $L_\theta^b$ is SLD of
 $\rho_\theta^b$. In this sense, we say that party $b$ cannot have any
 information about the parameter. Theorem 1 may be viewed as a no-imprint
 principle for QFI. It is more general than the no-imprint principle for
 quantum states~\cite{Mor1998}, where the restriction is keeping quantum
 states, stronger than keeping distinguishability.

 {\sl Broadcasting of QFI.}--In the broadcast scenario, the correlations
 between different parties are allowed. The broadcasting of quantum states
 always implies the broadcasting of QFI, but the converse is not true in
 general. In particular, in view of the no-broadcasting theorem for quantum
 states, namely, non-commuting states cannot be broadcast~\cite{Barnum1996}, QFI
 of a parametric family of commuting states can always be broadcast. On the
 other hand, since the previous no-broadcasting theorems show strong relevance
 with non-commutativity, one may ask whether the relation between
 non-commutativity and no-broadcasting is universal or not. Here, we show that
 QFI in some non-commuting quantum states can also be broadcast.

 First, we consider an illustrative example. The family of equatorial states
 $\{\sigma_\theta = \proj{\psi_\theta}\}_{\theta \in [0, 2\pi)}$ with
 $\ket{\psi_\theta} = \frac{1}{ \sqrt{2}} (\ket{0} + e^{i\theta}\ket{1})$ is
 apparently not a commuting family, and cannot be broadcast or cloned. Its
 approximate cloning has a remarkable application in quantum cryptography,
 through the link between cloning and eavesdropping~\cite{Drub2000,
   Scarani2005}. However, QFI of $\{\sigma _\theta\}$ can be broadcast: A
 Hadamard gate, followed by a controlled-NOT operation taking the input qubit
 as the source qubit and another qubit prepared in $|0\rangle $ as the target
 qubit, yields the states
\begin{equation*}
  \ket{\Psi_\theta ^{ab}}
  = e^{i\frac{\theta}{2}}
  \Big (\cos\frac{\theta}{2}\ket{00}-i\sin\frac{\theta}{2}\ket{11} \Big ),
\end{equation*}
from which we obtain $ F(\sigma_\theta)= F(\rho_\theta^a) =
F(\rho_\theta^b) = 1, $ with $\rho _\theta =|\Psi ^{ab}_\theta
\rangle \langle \Psi ^{ab}_\theta |$ and reduced states
$\rho_\theta^a$ and $\rho_\theta^b.$

It is of basic importance to determine for what kinds of parametric states,
QFI therein can be broadcast. We first have the following observations. (i)
QFI at a fixed parameter point, e.g., $\theta=\theta_0$, can be broadcast to
any number of parties, through the broadcasting of the outcomes of an optimal
distinguishing measurement $M_{\theta_0}=\{M_{j|\theta_0}\},$ which depends on
$\theta_0$. In other words, via the channel
\begin{equation}
  \label{eq:outcomes_broadcast_channel}
  \sigma_\theta \mapsto \rho_\theta =
  \sum_{j}\tr  \sigma_\theta M_{j|\theta _0}  \cdot (\ket{j}\bra{j} ) ^{\otimes n}
\end{equation}
we have $F(\rho_\theta^{(k)})=F(\sigma_\theta)$ at $\theta=\theta _0$, where
$\rho_\theta^{(k)}$ is the reduced states for party $k$. (ii) Unless the
measurement $M_{\theta_0}$ in Eq.~(\ref{eq:outcomes_broadcast_channel}) is
also optimal at other parameter points, the broadcasting channel
(\ref{eq:outcomes_broadcast_channel}) cannot be used to broadcast QFI at those
parameter points. These observations indicate that the broadcasting of QFI is
relevant to the dependence of the optimal distinguishing measurements on the
parameter. This motivates us to introduce the following definitions: We say
that QFI of $\{\sigma_\theta\}_{\theta\in\Theta}$ can be infinitely broadcast,
if it can be broadcast to any number of parties. We say that QFI of $\{\sigma
_\theta\}_{\theta\in \Theta}$ is uniform if there exists a single measurement,
which is optimal for distinguishing any $\theta \in \Theta,$ i.e., achieving
the equality in $F(\rho _\theta )=\max _M F(\rho _\theta |M).$ Now our main
result, which is reminiscent that a bipartite state is infinitely symmetric
extendible if and only if it is separable~\cite{SE1, SE3}, may be stated as
follows.

{\bf Theorem } 2. {\sl QFI of a parametric family of states
  $\{\sigma_\theta\}_{\theta \in \Theta}$ can be infinitely broadcast if and
  only if it is uniform.}

 First, it is clear that if QFI is uniform,
then it can be infinitely broadcast by the channel given by
Eq.~(\ref{eq:outcomes_broadcast_channel}). To establish the
converse, we  assume without loss of generality that the parameter
domain $\Theta$ is the unit interval $[0,1]$. Consider the states
$\{ \sigma_{i/n}\}_{i=0, 1, \cdots, n},$ we first show that if the
statistical distance between two neighboring states can be broadcast
to $m$ parties in the sense that  there exists a channel $\Lambda
:\sigma_{i/n}\mapsto \rho_{i/n} \in S(H^{\otimes m}) $ such that for
all $i$ and $k$,
\begin{equation*}
  \label{eq:SD_broadcasting}
  d_{\rm S} (\sigma_{i/n},\sigma_{(i+1)/n})
  =   d _{\rm S}(\rho_{i/n},\rho_{(i+1)/n})
  = d _{\rm S}(\rho_{i/n}^{(k)},\rho_{(i+1)/n}^{(k)}),
\end{equation*}
then  for every $m+1$ consecutive states in $\{\sigma
_{i/n}\}_{i=0,1,\cdots,
  n}$ there exists a single optimal measurement to distinguish
the neighboring states.  Here
$\rho_{i/n}^{(k)}$ is the reduced state for party $k$.  To prove
this, let us first consider the case $m=2$ (two parties $a$ and
$b$). The
 broadcasting condition, $
  d_{\rm S}(\rho_{i/n},\rho_{(i+1)/n})
= d_{\rm S}(\rho_{i/n}^{(k)},\rho_{(i+1)/n}^{(k)}), \ k=a, b$,
implies that there exist local optimal distinguishing measurements
$M^a_{i,i+1}$ and $M^b_{i,i+1}$ for the two neighboring states
$\rho_{i/n}$ and $\rho_{(i+1)/n}$. Moreover, $M^a_{i,i+1} \otimes
M^b_{i+1,i+2} $ is an optimal distinguishing measurement for both
the pairs of neighboring states ($\rho_{i/n},\rho_{(i+1)/n}$) and
($\rho_{(i+1)/n},\rho_{(i+2)/n}$), since the measurement-induced
probability distribution for party $a$ gives the largest possible
distinction between states $\rho_{i/n}$ and $\rho_{(i+1)/n}$, while
the one with respect to party $b$ gives the largest possible
distinguishability  between $\rho_{(i+1)/n}$ and $\rho_{(i+2)/n}$,
as illustrated in Fig.~2. In the meantime, note that if $M=\{M_j\}$
is an optimal distinguishing measurement for $\Lambda (\sigma_1)$
and $\Lambda (\sigma_2)$, then $ \Lambda ^\dagger(M):=\{\Lambda
^\dagger(M_j)\} $ is an optimal distinguishing measurement for
$\sigma_1$ and $\sigma_2.$ This is because performing $\Lambda
^\dagger(M)$ on $\sigma$ and performing $M$ on $\Lambda (\sigma)$
yield the same probability distribution and thus the same
distinguishability. Consequently, $ \Lambda ^\dagger
(M^a_{i,i+1}\otimes M^b_{i+1,i+2}) $ is an optimal distinguishing
measurement for both ($\sigma_{i/n},\sigma_{(i+1)/n}$) and
($\sigma_{(i+1)/n},\sigma_{(i+2)/n}$). This process  can be
straightforwardly extended to any $m$ parties, and leads to the
conclusion that if the statistical distance can be broadcast to $m$
parties, then there exists a single optimal distinguishing
measurement for every $m+1$ consecutive states. So the
infinite-broadcasting of the statistical distance implies a single
optimal distinguishing measurement  for all states. When $n\to
+\infty$, the statistical distance is proportional to QFI, and we
obtain the desired result.

\begin{figure} \center{
\includegraphics[height=3cm]{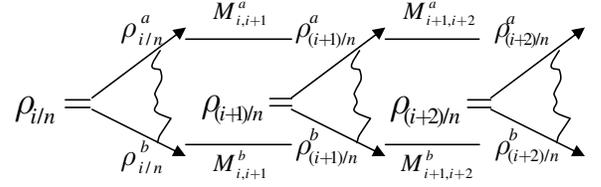}
\caption {\label{fig:2} {\small
    Broadcasting of the statistical distance vs. the uniformness of QFI.
    $M^a_{i,i+1}$ and $M_{i,i+1}^b$ are local optimal distinguishing
    measurements for the  local neighboring states
    $(\rho^a_{i/n},\rho^a_{(i+1)/n})$ and $(\rho^b_{i/n},\rho ^b_{(i+1)/n}),$
    respectively, $M_{i,i+1}^a \otimes M_{i+1,i+2}^b$ is an optimal global
    distinguishing measurement for both $(\rho_{i/n},\rho_{(i+1)/n})$ and
    $(\rho_{(i+1)/n},  \rho_{(i+2)/n})$ due to the broadcasting condition.
    When $n$, which is the number of divisions for the parametric family
    states, tends to infinite, the statistical distance between neighboring
    states tends to QFI, and thus broadcasting infinitesimal statistical
    distance amounts to broadcasting QFI.
}}}
\end{figure}

It should be emphasized that for finite broadcasting, the
uniformness of QFI is not necessary for the broadcasting of QFI. In
order to elucidate this, we first identify all states arising from
broadcasting of QFI.  We say that $\rho_\theta \in S(H^{\otimes n})$
is a family of $n$-partite QFI-broadcast states in the parameter
domain $\Theta$, if $F(\rho_\theta^{(k)}) = F(\rho_\theta)$ for
every $\theta \in \Theta$ and every $k$. In what follows, we will
take the notational convenience such as $ X^a Y^{ab} = (X^a \otimes
{\bf 1}^b) Y^{ab}.$

{\bf Theorem } 3. {\sl $\rho_\theta \in S (H^{\otimes n})$ is a family of
  $n$-partite QFI-broadcast states if and only if any SLD for the reduced
  states of any party is also an SLD for $\rho_\theta$.}

The sufficiency readily follows from the calculation of QFI. Indeed, if
$L_\theta^{(k)},$ an SLD for $\rho_\theta^{(k)}$, is also an SLD for
$\rho_\theta$, then
\begin{equation*}
  F(\rho_\theta)
  = \tr \rho_\theta  L_\theta^{(k)2}
  = \tr \rho_\theta^{(k)} L_\theta^{(k)2}
  = F(\rho_\theta^{(k)}).
\end{equation*}

To prove the necessity, we make use of the optimal distinguishing
measurements. The condition $F(\rho_\theta) = F(\rho_\theta^{(k)})$ implies
that there exists a local optimal measurement
$M^{(k)}_\theta=\{M^{(k)}_{j|\theta} \}$ for distinguishing $\theta$. On the
other hand, it is known that $M_\theta=\{M_{j|\theta}\}$ is an optimal
distinguishing measurement for $\theta$ if and only if~\cite{Braunstein1994}
\begin{equation}
  \label{eq:optimal_POVM_condition}
M_{j|\theta}^{1/2} L_\theta \rho_\theta^{1/2}  = u_{j|\theta}
M_{j|\theta}^{1/2} \rho_\theta^{1/2},
  \quad \forall \ j,
\end{equation}
where $ u_{j|\theta}:=\partial_\theta {\rm log} \tr \rho_{\theta} M_{j|\theta}
$ when $ {\rm } \rho_{\theta} M_{j|\theta} $ is nonzero, and vanishes
otherwise. Equation (\ref{eq:optimal_POVM_condition}) implies that $ L_\theta
\rho_\theta = \Big( \sum_j u_{j|\theta} M_{j|\theta} \Big) \rho_\theta, $
which in turn indicates that $ \sum_j u_{j|\theta} M_{j|\theta} $ is an SLD
for $\rho_\theta$. Because $M_\theta^{(k)}$ is an optimal distinguishing
measurement not only for $\rho_\theta^{(k)}$ but also for $\rho_\theta$,
 \begin{equation}
   \label{eq:SLD_POVM}
   L_\theta^{(k)} :=
   \sum_j \partial_\theta {\rm log}
   \Big ( \tr \rho_\theta M_{j|\theta}^{(k)} \Big ) \cdot M^{(k)}_{j|\theta}
 \end{equation}
  is an SLD for both $\rho_\theta^{(k)}$ and $\rho_\theta$.
Moreover, every SLD for $\rho_\theta^{(k)}$ can be expressed in the
form of Eq.~(\ref{eq:SLD_POVM}), e.g., taking $M_{j|\theta}$ as the
eigenstates of the SLD. Thus, for QFI-broadcast states
 $\rho_\theta$, every  SLD for $\rho^{(k)}_\theta$ is also an SLD for
 $\rho_\theta$.

 From Theorem 3, we can derive an important property of QFI-broadcast states:
 {\sl $\big[\rho^{(k)}_\theta,\partial_\theta\rho^{(k)}_{\theta}\big]=0$ for
   the reduced states of any QFI-broadcast state $\rho_\theta$.}

 To establish this, without loss of generality, we consider two parties $a$
 and $b$. According to Theorem 3, $L_\theta^a$ and $L_\theta^b$ are both SLD
 for $\rho_\theta $. Due to the uniqueness of the product of SLD and the
 state~\cite{Fujiwara1999}, we have $L_\theta^a\rho_\theta = L_\theta^b
 \rho_\theta $, which implies $ L_\theta^a \rho_\theta^a = \tr_b L_\theta^a
 \rho_\theta = \tr_b L_\theta^b \rho_\theta .$ Similarly, we have $
 \rho_\theta^a L_\theta^a = \tr_b \rho_\theta L_\theta^a = \tr_b \rho_\theta
 L_\theta^b.$ From the cyclic property of trace, $ \tr_b L_\theta^b
 \rho_\theta = \tr_b \rho_\theta L_\theta^b, $ we get $L_\theta^a
 \rho_\theta^a = \rho_\theta^a L_\theta^a $.

 Now we have a better understanding of the relation between non-commutativity
 and no-broadcasting. QFI-broadcasting is necessary, but not sufficient, for
 the broadcasting of quantum states. It is the requirement of the broadcasting
 of quantum states, i.e., $\rho_{\theta}^{(k)}=\sigma_{\theta}$, that
 transfers the commutativity from reduced states of the output states to the
 input states. This can also be seen from the original proof of the
 no-broadcasting theorem for quantum states~\cite{Barnum1996}.

 The neighboring commutativity, $
 \big[\rho^{(k)}_\theta, \partial_\theta\rho^{(k)}_{\theta} \big]=0$, most
 likely but not necessarily leads to the (pairwise) commutativity. If there
 exists a party for which the reduced states $\rho_\theta^{(k)}$ of the
 QFI-broadcast states are pairwise commuting, then the common eigenbases, with
 which all $\rho_\theta^{(k)}$ with $\theta\in\Theta$ are simultaneously
 diagonalized, make up a single local optimal distinguishing measurement for
 $\{\rho_\theta\}_{\theta \in \Theta}$. Thus, QFI of $\{\rho_\theta\}_{\theta
   \in \Theta}$ is uniform, which in turn implies the uniformity of QFI of
 input states $\{\sigma_\theta\}_{\theta\in\Theta}$, and hence their infinite
 broadcastability.

 To show that the uniformness of QFI is not necessary for {\sl finite}
 broadcasting of QFI, we explicitly construct a family of bipartite
 QFI-broadcast states whose QFI is not uniform. Without requiring the
 smoothness of the parametrization, such a family of two-qubit states can be
 constructed as $\sigma_\theta = \proj{\Psi_\theta}$ with
$$
\ket{\Psi_\theta} =
\begin{cases}
  \ket{\psi_{yy}(-\theta)}, &\mbox{if } \theta \in [-\pi/2,-\pi/4],  \\
  \ket{\psi_{zz}(\theta)}, &\mbox{if } \theta \in (-\pi/4,\pi/4]  ,\\
  \ket{\psi_{xx}(\theta)}, &\mbox{if } \theta \in (\pi/4,\pi/2],
\end{cases}
$$
where $\ket{\psi_{ii}(\theta)} = \cos\theta\ket{i}\otimes\ket{i} +
\sin\theta\ket{\bar{i}}\otimes\ket{\bar{i}}$ for $i=x$, $y$ and $z$,
$\ket{x}$, $\ket{y}$ and $\ket{z}$ are the eigenstates of the corresponding
Pauli matrices with the eigenvalue $1$, while $\ket{\bar{x}}, \ket{\bar{y}}$
and $\ket{\bar{z}}$ are those with eigenvalue $-1$. QFI of
$\{\sigma_\theta\}_{\theta\in [-\pi/2,\pi/2]}$ is not uniform, since there
does not exist a single optimal distinguishing measurement for the whole
domain, however, it can be broadcast by distributing the two qubits to two
parties directly, since $\sigma_\theta$ are themselves bipartite QFI-broadcast
states.

{\sl Discussion.}--By measuring the statistical distinguishability of a
parameter in terms of QFI, we have established a no-cloning theorem for QFI
and a no-infinite-broadcasting theorem for non-uniform QFI. The results extend
the no-broadcasting theorem for quantum states ~\cite{Barnum1996}, and shed
new lights on quantum communication and quantum metrology. It is also
interesting to note that QFI-broadcasting is complementary to the
environment-assisted precision measurement~\cite{Goldstein2011}, whose essence
is the concentration of QFI from the system-environment entirety into the
system.

The broadcasting and manipulation of QFI is of broad significance for quantum
information processing, opens new research directions, and indicates an
alternative way in understanding the boundary and link between classical and
quantum information.

Firstly, in practice, many important quantities are represented by parameters,
which in turn are physically encoded in quantum states, and thus the
estimation or measurement of the parameters, rather than the quantum states
themselves, are of practical relevance. It is much costly, and actually not
necessary, to gain complete information about the quantum states. QFI
information is a fundamental ingredient in such a scenario. For example, the
phase estimation is an extremely important issue. It is the relative phase in
superposition or coherence, rather than the superposed quantum states
themselves that needs to be estimated or measured.

Secondly, in quantum foundations, a crucial and long standing problem is to
depict the boundary between classical and quantum information, which actually
lies in the heart of quantum decoherence and quantum measurement. This may
first require our deeper understanding of quantities lying between classical
and quantum nature. QFI is precisely such an intrinsic and hybrid character.
In fact, QFI represents a kind of information intermediate between the
classical parameters (which can be freely cloned and broadcast) and complete
quantum information (represented by quantum states, but cannot be broadcast in
general). Thus the understanding of QFI may serves as a bridge connecting our
theories between classical and quantum and in particular may open a new avenue
in attacking the manifestation of classicality from quantum substrate.

Thirdly, QFI broadcasting has immediate implications in some quantum metrology
problems, where quantum sensors and measurement apparatus are connected by
noise channels. In order to overcome the loss and increase the receiving
efficiency, one can utilize the broadcasting technology to distribute
information. However, broadcasting quantum states is not only very restricted,
but also unnecessary in many cases. Broadcasting QFI relaxes the restrictions
to a great extent, shows much more flexibility, and keeps the essence of
information. More importantly, equatorial states, which are the most often
used ones in quantum precision measurement, can be QFI-broadcast.

Finally, we mention the important subject of protection and transmission of
quantum Fisher information in theoretical and experimental quantum parameter
estimation, as opposed to the protection and transmission of quantum states,
which are much more costly and complicated. For example, decoherence-free
subspaces for phase estimation still focus on the protection of quantum states
\cite{Decoherencefree}. The situation will be quite different if we consider
the protection of QFI instead. Our work can help to develop novel strategies
for robust quantum enhanced phase estimation against noise.

\begin{acknowledgments}
  Lu thanks P.-Q. Jin for helpful communication. This work was supported by
  National Research Foundation and Ministry of Education, Singapore, Grant No.
  WBS: R-710-000-008-271, the NFRPC, Grant No. 2012CB921602, NSFC, Grant Nos.
  11025527, 10935010, and 11005027, and Program for HNUEYT, Grant No.
  2011-01-011.
\end{acknowledgments}

\end{document}